\documentclass[twocolumn,prl,secnumarabic,amssymb,amsmath,nobibnotes,aps,showpacs]{revtex4-1}

\usepackage{graphics}
\usepackage{graphicx}
\usepackage{braket}
\usepackage{float}
\usepackage{multirow}

\usepackage[usenames,dvipsnames]{color}

\begin{document}

\title{Sub-millikelvin dipolar molecules in a radio-frequency magneto-optical trap}

\author{E. B. Norrgard$^{1}$, D. J. McCarron$^{1}$, M. H. Steinecker$^{1}$,
M. R. Tarbutt$^{2}$, and D. DeMille$^{1}$}

\affiliation{$^{1}$Department of Physics, Yale University, P.O. Box 208120, New Haven, Connecticut 06520, USA}
\affiliation{$^{2}$Centre for Cold Matter, Blackett Laboratory, Imperial College London, Prince Consort Road, London SW7 2AZ UK}

\begin{abstract}
We demonstrate a scheme for magneto-optically trapping strontium monofluoride (SrF) molecules at temperatures one order of magnitude lower and phase space densities three orders of magnitude higher than obtained previously with laser-cooled molecules.  In our trap, optical dark states are destabilized by rapidly and synchronously reversing the trapping laser polarizations and the applied magnetic field gradient.  The number of molecules and trap lifetime are also significantly improved from previous work by loading the trap with high laser power and then reducing the power for long-term trapping.  With this procedure, temperatures as low as 400 $\mu$K are achieved.
\end{abstract}


\maketitle

% Introduction
Recently, there has been great interest and advancement in producing samples of cold and ultracold ($T \textless$  1\,mK) polar molecules \cite{Carr2009}.  The rich internal structure of molecules naturally lends itself to diverse and exciting applications including ultracold chemistry \cite{Krems2008}, precision measurements \cite{Hunter2012, Tarbutt2013}, and quantum simulation \cite{Micheli2006}.  A number of indirect \cite{Ni2008, Danzl2010, Aikawa2010, Shimasaki2015} and direct \cite{Zeppenfeld2012, Stuhl2012, Lu2014, Zhelyazkova2014} methods for obtaining ultracold molecules have been demonstrated or are under development.  Recently, we have demonstrated use of a magneto-optical trap (MOT) \cite{Barry2014,McCarron2015} to directly cool and trap laser-slowed \cite{Barry2012} SrF molecules.  A MOT provides simultaneous confinement and cooling, making it the starting point of nearly all experiments with ultracold atoms; the MOT is similarly promising for use with molecules.  In this Letter we demonstrate and characterize a scheme for producing a molecular MOT that yields much lower temperature and much higher phase space density than in our previous work.

Most atomic MOTs use a type-I level structure \cite{Prentiss1988} ($F$\,$\rightarrow$\,$F'$\,=\,$F+1$, where $F$ is the total angular momentum and the prime indicates the excited state), where all ground states are bright states (here, defined as states addressed by a laser beam that provides a radiative confining force).
In molecules, rotational structure requires use of a type-II configuration ($F$\,$\rightarrow$\,$F'$\,=\,$F$ or $F$\,$\rightarrow$\,$F'$\,=\,$F-1$), which has ground state sublevels not coupled to a confining laser (dark states) for any fixed polarization.
The dark state population must be rapidly returned to bright states to allow significant cooling and confinement. Typically, MOTs based on type-II structures are observed to yield weaker confining forces and higher temperatures than those using a type-I structure \cite{Oien1997, Tiwari2008}.

\begin{figure}[!b]
\centering
\includegraphics[width=\linewidth]{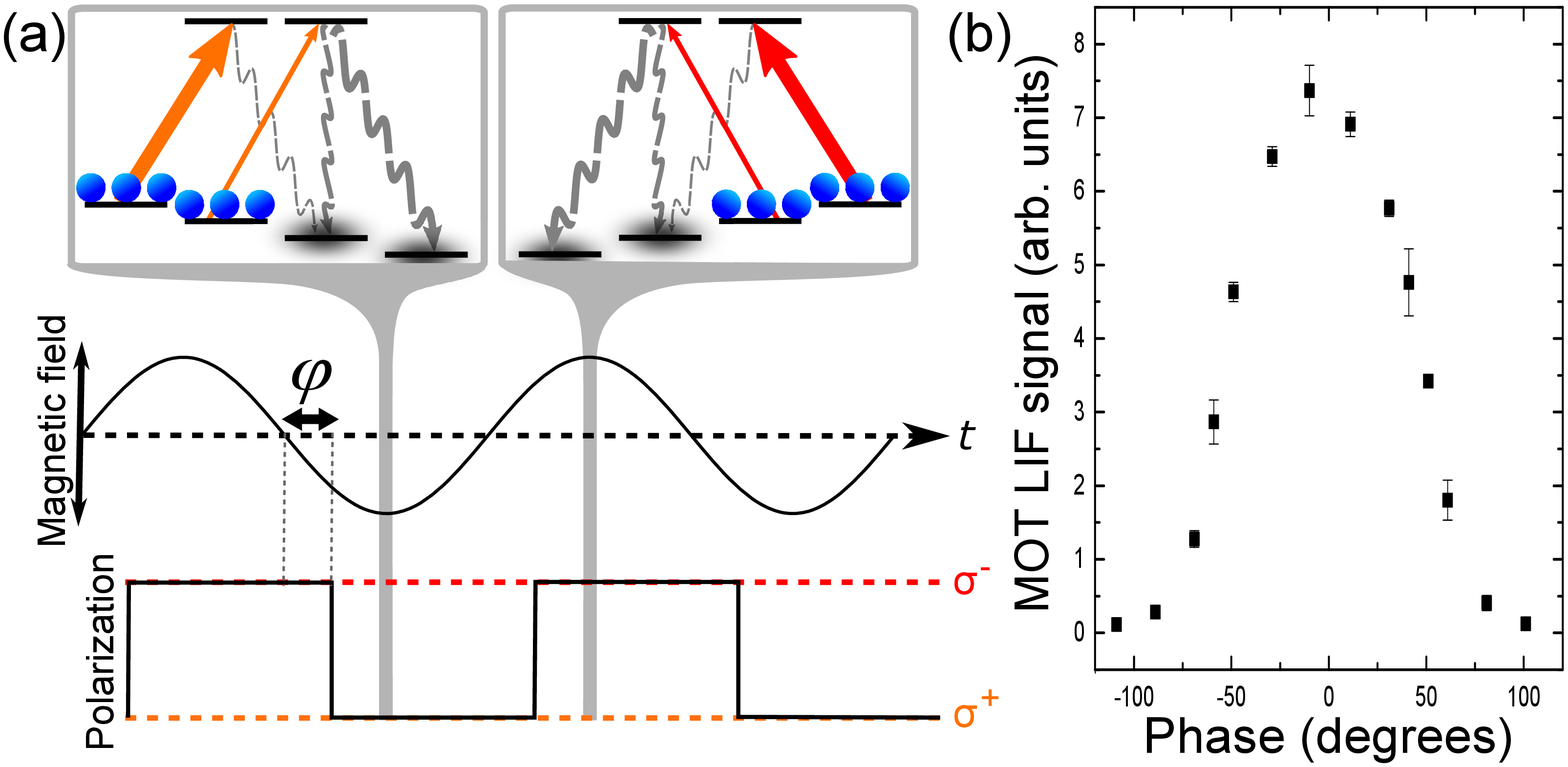}
        \caption{(Color online) (a) RF MOT trapping concept. Excitation by a confining laser (solid arrows) may lead to decay (dashed arrows) into dark states. The $B$-field and polarization rapidly reverse such that dark states become bright states; this leads to increased confinement and cooling. \mbox{(b) LIF} from trapped molecules vs relative phase $\phi$ of $B$-field and laser polarization.  Molecules are only trapped for $|\phi|$\,$\lesssim$\,$90^{\circ}$.}
        \label{fig:FigA}
\end{figure}

In this paper we demonstrate a molecular MOT which destabilizes dark states using time-varying fields.
A 3D confining force is produced by diabatically transferring molecules from dark states to bright states; this is accomplished by rapidly and synchronously reversing the $B$-field and the trapping light polarization \cite{Hummon2013} (Fig.\,\ref{fig:FigA}a).  If the switching is done at a frequency $f_{\rm{MOT}}$\,$\gtrsim$\,$R_{\rm{sc}}$, where $R_{\rm{sc}}$ is the molecules' photon scattering rate, then the molecules spend a reduced amount of time in dark states and should experience a correspondingly larger force.
For a two-level system the maximum value of $R_{\rm{sc}}$ is $\Gamma/2$, where the excited state decay rate $\Gamma$ is typically on the order $\Gamma$\,$\sim$\,$10^7$\,s$^{-1}$ \cite{DiRosa2004} ($\Gamma$\,=\,$2\pi$$\times$7\,MHz for SrF).  Hence we refer to the regime \mbox{$f_{\rm{MOT}}$\,$\gtrsim$\,$R_{\rm{sc}}$} as the radio-frequency (RF) MOT; we distinguish it from the regime \mbox{0\,$<$\,$f_{\rm{MOT}}$\,$\lesssim$\,$R_{\rm{sc}}$}, where dark states are destabilized predominantly through other means, which we refer to as the AC MOT.  The RF MOT principle was previously used to apply 2D transverse compression on a beam of YO molecules \cite{Hummon2013}, and an AC MOT was used to trap K atoms \cite{Harvey2008}.

 In previous work \cite{Barry2014,McCarron2015}, we demonstrated MOTs of SrF with static polarizations and $B$-field.  Much of this paper is devoted to a comparison between these MOTs (which we review here briefly) and the RF MOT.
 The ground state of SrF has resolved spin-rotation/hyperfine (SR/HF) structure; each of these sublevels must be addressed by a separate optical frequency and polarization to achieve reasonable scattering forces.
 In our original static molecular MOT \cite{Barry2014} (referred to here as the DC MOT), the polarizations of the individual frequency components \cite{SMfigure} were chosen so that the states Zeeman-shifted closest to resonance were bright states. Surprisingly, this scheme does not maximize the restoring force \cite{Tarbutt2015}, and the confinement was very weak compared to typical atomic MOTs. In subsequent work \cite{McCarron2015}, the polarizations of the components were chosen so that each, treated independently, provided a restoring force \cite{Tarbutt2015,TarbuttII2015}. This scheme, referred to as the DC* MOT, gave increased confining forces but significantly higher temperatures. In the RF MOT, the polarizations are the same as in the original DC MOT \cite{SMfigure}.

  In this work, we show that the RF MOT makes it possible to obtain much higher phase space density than in the DC or DC* MOTs: it provides significant confinement and long lifetime even at low scattering rates, where the lowest temperatures are achieved.  Some qualitative explanation of this behavior will be provided by an analytical model of the system.

% Experimental Setup and Procedure
Much of the experimental scheme has been described elsewhere \cite{Shuman2010,Barry2011,Barry2012,Barry2014,McCarron2015}.  Briefly, SrF molecules from a cryogenic beam source \cite{Patterson2007,Hutzler2012,Barry2011} are slowed by lasers $\mathcal{L}_{00}$, $\mathcal{L}_{10}$, and $\mathcal{L}_{21}$ \cite{Barry2012}, where $\mathcal{L}_{vv'}$ denotes a laser tuned to the $\ket{X^2\Sigma,N=1,v}\rightarrow \ket{A^2\Pi_{1/2},J=1/2,P=+,v'}$ transition (where $P$ is the parity).  Slow molecules are captured by the MOT, which contains the additional vibrational repump laser $\mathcal{L}_{32}$; due to SrF's highly diagonal matrix of Franck-Condon factors \cite{DiRosa2004}, these four lasers allow each molecule to scatter $\sim$3$\times10^6$ photons before decaying to uncoupled vibrational levels ($v$\,$\geq$\,4). Laser $\mathcal{L}_{00}^{N=3}$ addresses decay to $\ket{v=0,N=3}$ caused by off-resonant excitation from the optical cycle to $J'$\,=\,3/2 excited states  \cite{McCarron2015}.  SR/HF structure is addressed by adding RF sidebands to each of these MOT lasers.  Because one SR/HF level requires a different polarization for trapping \cite{Barry2014,Tarbutt2015}, an additional single-frequency trapping laser $\mathcal{L}_{00}^{\dagger}$ with polarization opposite to $\mathcal{L}_{00}$ and tuned closer than the closest sideband of $\mathcal{L}_{00}$ to this line allows for a better approximation to the ideal polarization scheme \cite{Barry2014,McCarron2015}.  An acousto-optic modulator (AOM) allows the power of main cycling lasers $\mathcal{L}_{00}$ and $\mathcal{L}_{00}^{\dagger}$ to be rapidly changed.  Losses from additional optics reduce the maximum $\mathcal{L}_{00}$ and $\mathcal{L}_{00}^{\dagger}$ power to 80\,mW and 30\,mW, respectively, compared to 210\,mW and 50\,mW in our previous work \cite{Barry2014,McCarron2015}.  Trapped molecules are detected by imaging laser-induced fluorescence (LIF) from the main cycling transition onto a CCD camera.

We rapidly switch the polarization of the main cycling lasers using a Pockels cell.  The sinusoidally oscillating $B$-field gradient is produced by a pair of in-vacuum aluminum nitride (AlN) boards with direct bond copper coils on each side.  Variable capacitors external to the vacuum chamber form a parallel LC tank circuit with the coils resonant at frequency $f_{\rm{MOT}}$, and impedance match to 50 $\Omega$.  Adjusting the capacitors allows $f_{\rm{MOT}}$ to be easily varied from DC\,-\,3\,MHz (limited by the coils' self-capacitance) with loaded $Q\approx50$ for $f_{\rm{MOT}}>0.5$\,MHz.

 A peak current of 5\,A is required to produce the optimum RMS axial $B$-field gradient of 9\,G/cm.  Given our coil inductance of 40\,$\mu$H, this requires a peak voltage \mbox{$V_{\rm{pk}}$\,=\,4\,kV} when $f_{\rm{MOT}}$\,=\,3\,MHz.  We take care to minimize the $\mathcal{E}$-field produced by this voltage drop: the four coils (one on each side of each board) are wired in series, with the coils on the bottom board ordered first and last, and top board coils intermediate, leading to equal average potentials on the two boards.  Nevertheless the residual $\mathcal{E}$-field has observable effects on the trap lifetime (discussed below).

%Properties of Loading

Defining the time of the laser ablation pulse that initiates a molecular beam pulse as $t$\,=\,0, we find the maximum population occurs at $t$\,$\approx$\,70\,ms.  Fig.\,1b shows the LIF recorded for a 60\,ms exposure starting at $t$\,=\,70\,ms as a function of the relative phase $\phi$ of the $B$-field gradient and laser polarization.  As expected, the signal is greatest when $\phi$\,$\approx$\,0, and vanishes for $|\phi|$\,$\gtrsim$\,$90^{\circ}$. This clearly demonstrates trapping via the RF MOT mechanism.  For optimal trap loading, $\mathcal{L}_{00}$ and $\mathcal{L}_{00}^\dagger$ are detuned from a nominal resonant frequency (the frequency which produces maximum LIF when the lasers are applied orthogonal to the molecular beam in zero $B$-field) by about \mbox{$-9$\,MHz} and $-6$\,MHz, respectively.

\begin{figure}[!b]
\centering
\includegraphics[width= .8\linewidth]{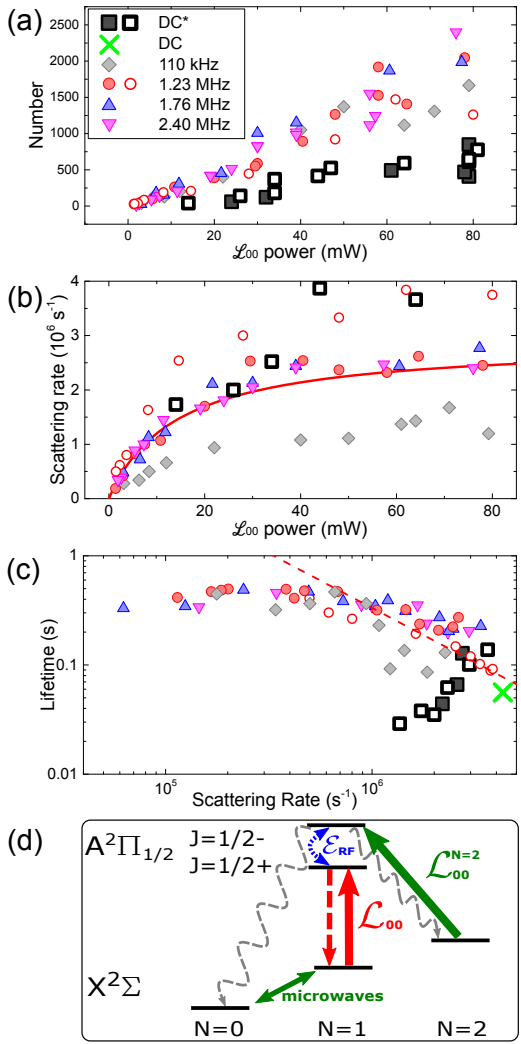}
        \caption{(Color online.  Marker legend in figure. Filled markers with microwaves applied, open markers without.) \mbox{(a) Number} of molecules loaded into the MOT vs $\mathcal{L}_{00}$ power.  The number increases with increasing $\mathcal{L}_{00}$ power in all cases.  (b)  Scattering rate vs $\mathcal{L}_{00}$ power. Solid line is a fit of the model described in the text to $f_{\rm{MOT}}$\,=\,1.23\,MHz data, with $\Delta$ set to $-2\pi$\,$\times$\,$9$\,MHz and fit values $\Gamma_{\rm{eff}} = 2\pi$\,$\times$\,$1.0(1)$\,MHz and $p_{\rm{sat}}$\,=\,$1.9(3)$\,mW.  $R_{\rm{sc}}$ is lower for $f_{\rm{MOT}}$\,=\,111\,kHz than for $f_{\rm{MOT}}$\,$>$\,1\,MHz.  $R_{\rm{sc}}$ increases in the absence of repump microwaves by $\sim$40$\%$.  \mbox{(c) Trap} lifetime $\tau_{\rm{MOT}}$ vs $R_{\rm{sc}}$.  $\tau_{\rm{MOT}}$ as long as 500\,ms are achieved for low $R_{\rm{sc}}$ in the RF MOT.  For higher $R_{\rm{sc}}$, $\tau_{\rm{MOT}}$\,$\propto$\,$1/R_{\rm{sc}}$ (dashed line is a $1/R_{\rm{sc}}$ trend line), and $\tau_{\rm{MOT}}$ is much shorter in the absence of repump microwaves.  For the DC* MOT, $\tau_{\rm{MOT}}$ decreases for lower $R_{\rm{sc}}$. (d) Parity mixing (curved blue) due to the coil-induced electric field $\mathcal{E}_{RF}$ causes loss (wiggly grey) from the main cycling transition (vertical red), which is repumped via $\mathcal{L}_{00}^{N=2}$ and resonant microwaves (diagonal green).}\label{fig:FigD}
\end{figure}

The number of trapped molecules is determined from $R_{\rm{sc}}$, MOT lifetime $\tau_{\rm{MOT}}$, and integrated LIF intensity from a 60\,ms camera exposure as in \cite{Barry2014}.
The peak trap population vs $\mathcal{L}_{00}$ power is plotted in Fig.\,$\ref{fig:FigD}$a (throughout this work, whenever $\mathcal{L}_{00}$ varies from full power, it is implicit that the $\mathcal{L}_{00}^{\dagger}$ power is changed by a proportional amount using the AOM).  The number of trapped molecules grows monotonically with the laser power.  Under optimal conditions, $>$$2000$ molecules are loaded into the RF MOT at full power, and there is no clear dependence on $f_{\rm{MOT}}$ in the range \mbox{1.23\,-\,2.40\,MHz}.  Operation at $f_{\rm{MOT}}=111$\,kHz was much more sensitive to MOT beam alignment than for $f_{\rm{MOT}}>1$ MHz, so we speculate the slightly lower number loaded may be attributed to drifting beam pointing.  We find that the RF MOT loads $\sim$$3\times$ more molecules than the DC* MOT.

The photon scattering rate $R_{\rm{sc}}$ (plotted vs $\mathcal{L}_{00}$ power in Fig.\,\ref{fig:FigD}b) is determined by comparing the measured decay rate of the LIF signal when $\mathcal{L}_{32}$ is rapidly shuttered vs when $\mathcal{L}_{32}$ is present, given the known branching ratio into $v=3$ ($\approx9.6\times10^{-6}$ \cite{Stienkemeier1993, Barry2013}).  A simple analytical rate model predicts that the steady-state scattering rate in the MOT has the same form as a two-level system \cite{Tarbutt2013}:
\begin{equation}
R_{sc} = \frac{\Gamma_{\rm{eff}}}{2} \frac{I_{00}/I_{\rm{sat,eff}}}{1 + I_{00}/I_{\rm{sat,eff}} + 4\Delta^{2}/\Gamma^{2}}.
\label{eq:Rsc}
\end{equation}
Here $I_{00}$ is the ${\cal L}_{00}$ intensity, $I_{\rm{sat,eff}}$ is an effective saturation intensity, $\Gamma_{\rm{eff}}/2$ is the maximum scattering rate, and $\Delta$ is the detuning from resonance. The model is discussed in more detail in the Supplemental Materials \cite{SMgeneral}, where expressions for $\Gamma_{\rm{eff}}$ and $I_{\rm{sat,eff}}$ in terms of known molecular parameters are also given. These expressions predict \mbox{$\Gamma_{\rm{eff}}$\,=\,$2\pi\times$\,1.7\,MHz} and $I_{\rm{sat,eff}}$ corresponding to saturation power $p_{\rm{sat}}$\,=\,2.6\,mW for our 7\,mm 1/$e^2$ radius MOT beams. The solid line in Fig.\,2b shows a fit of Eq.\,(\ref{eq:Rsc}) to the data for $f_{\rm{MOT}}$\,=\,1.23\,MHz. We fix $\Delta=-2\pi\times$\,9\,MHz, and find the best fit parameters to be \mbox{$\Gamma_{\rm{eff}}$\,=\,$2\pi\times$\,1.0(1)\,MHz}, about 40\% smaller than the predicted value, and $p_{\rm{sat}}$\,=\,1.9(3)\,mW, about 25\% smaller than predicted. We see that, despite the complexity of the MOT, this simple model predicts the scattering rate quite well. However, we are unable to quantitatively explain the much lower $R_{\rm{sc}}$ observed for $f_{\rm{MOT}}$\,=\,111\,kHz, which is well into the AC MOT regime for all but the very lowest $R_{\rm{sc}}$ explored here.

In simplifying limits, other well-known equations approximating trap properties, such as the spring constant $\kappa$ and temperature $T$, may be expressed in terms of either $R_{\rm{sc}}$ or  $I_{00}/I_{\rm{sat,eff}}$.   Because $R_{\rm{sc}}$ may be easily measured in the lab, and the expected dependence of the trap properties is simpler when expressed in terms of $R_{\rm{sc}}$, we choose to plot the remainder of our data as function of $R_{\rm{sc}}$ rather than power or intensity.

The MOT lifetime $\tau_{\rm{MOT}}$ is measured by monitoring LIF as a function of time (Fig.\,\ref{fig:FigD}c). RF $\mathcal{E}$-fields caused by the large RF voltage applied to the MOT coils mix the excited $\ket{J=1/2, P=\pm}$ states and lead to unwanted branching to the $N$\,$=$\,$0,2$ ground states.  In the absence of a mechanism to couple these states to the main cycling transition, this leads to relatively short trap lifetimes.  As shown in Fig.\,2d, we mitigate this loss with a $\sim$5\,mW laser $\mathcal{L}_{00}^{N=2}$ (which pumps $N$\,=\,2 into $N$\,=\,0 via $\ket{A^2\Pi_{1/2}, J=1/2, P=-}$) plus microwaves resonant with both $\ket{N=0,J=1/2,F=0,1} \leftrightarrow \ket{N=1,J=1/2,F=1,0}$ transitions.  For intermediate scattering rates ($R_{\rm{sc}}$\,$\sim$2\,-\,8$\times10^5$\,s$^{-1}$) with these additional fields applied, $\tau_{\rm{MOT}}$ is independent of $f_{\rm{MOT}}$ and is especially long ($\sim$500\,ms).  Despite decreasing the maximum $R_{\rm{sc}}$ by $\sim$40$\%$ (roughly the expected amount due to additional levels coupled to the cycling transition \cite{SMgeneral}, see Fig.\,2b), microwaves are found to not dramatically affect other trap properties.  For the highest scattering rates, the trap loss rate $R_{\rm{loss}}=1/\tau_{\rm{MOT}}$ is proportional to $R_{\rm{sc}}$, indicative of a scattering-related loss mechanism whose origin is not currently understood.  In contrast to the RF MOT, the DC* MOT lifetime decreases sharply for lower scattering rates, rendering it incapable of trapping at $R_{\rm{sc}} \lesssim 10^6$\,s$^{-1}$.  The DC MOT was found to be even more short-lived at any reduced power; the single points in Figs.\,2 and 3 were taken from \cite{Barry2014}.

\begin{figure*}[t!]
\centering
\includegraphics[width= \textwidth]{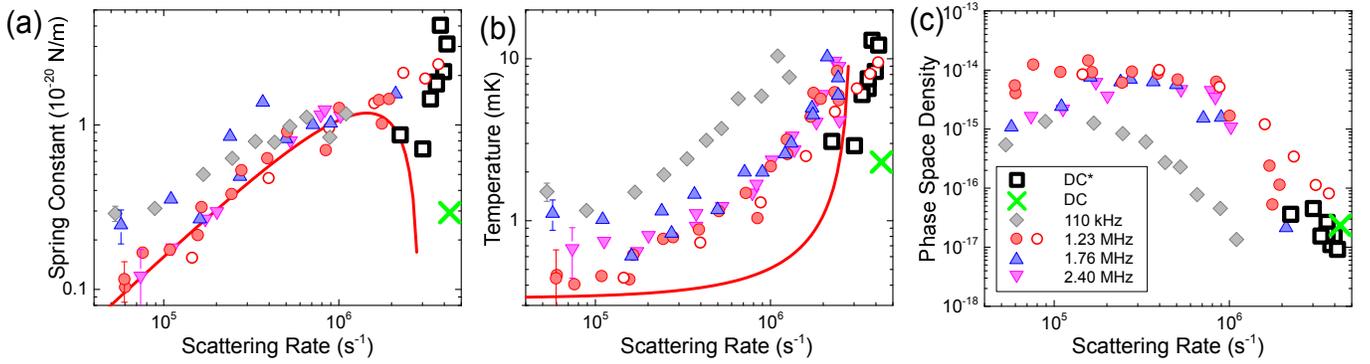}
        \caption{(Color online.  Marker legend in figure. Filled markers with microwaves applied, open markers without. Red line analytical model fit to $f_{\rm{MOT}}$\,=\,1.23\,MHz data.) Measured (a) spring constant $\bar{\kappa}$ and (b) temperature $\bar{T}$ vs scattering rate $R_{\rm{sc}}$.  The maximum $\bar{\kappa}$ for the DC* MOT is $\sim$2$\times$ that of the RF MOT, but decreases much more rapidly for lower $R_{\rm{sc}}$.  The analytical model, with $\Gamma_{\rm{eff}}$ set to the fit value from $R_{{sc}}$ vs $I_{00}$, and $B$-field gradient $\partial B_z/\partial z$ and detuning $\Delta$ set to experimental values, has $\mu_{\rm{eff}}$ as the only free parameter. $\bar{T}$ decreases with falling $R_{\rm{sc}}$, but at a slower rate than predicted by the analytical model with no free parameters.  $\bar{T}$ is lower for $f_{\rm{MOT}}$\,$>$\,1\,MHz than for $f_{\rm{MOT}}$\,=\,\,111 kHz.  The model correctly predicts $\bar{T}\sim T_D$ for low $R_{\rm{sc}}$. The spread in values for a given MOT configuration are indicative of variations between measurements made on the same day, while the representative error bars are standard errors of measurements taken on multiple days under similar conditions. (c) Phase space density $\rho$ vs $R_{\rm{sc}}$ after a linear power ramp from full power.  Ramping increases $\rho$ by up to 2.5 orders of magnitude in the RF MOT. In the DC* MOT, $\rho$ increases only slightly when $R_{\rm{sc}}$ is lowered.}\label{fig:FigC}
\end{figure*}

While the MOT loads the most molecules at full laser power, we observe other trap properties improve as the power of the main cycling lasers is reduced.   We therefore employ a loading stage at full laser power to capture the largest possible sample, then reduce the $\mathcal{L}_{00}$ and $\mathcal{L}_{00}^{\dagger}$ powers.  By linearly decreasing the power from its full initial value to a lower final value over a time $t_{\rm{ramp}}$\,$>$\,30\,ms, the fraction of trapped molecules remaining is near unity, even when the power is reduced by $99.5$\,\%.  For all data in Fig.\,3, we use $t_{\rm{ramp}}$\,=\,50\,ms.

%RF MOT Properties

	The radial and axial temperatures $T_{\rho}$ and $T_z$ and spring constants $\kappa_\rho$ and $\kappa_z$ are measured by observing the free expansion of molecules released from the MOT.  The trapping potential is removed by turning off the $B$-field and the main cycling lasers for a variable duration $t_{\rm{TOF}}$, during which time the molecules ballistically expand.  The main cycling lasers are then restored, and the molecule cloud is imaged for 2\,ms.

	The geometric mean spring constant $\bar{\kappa} \equiv \kappa_\rho^{2/3} \kappa_z^{1/3}$ is shown as a function of $R_{\rm{sc}}$ in Fig.\,$\ref{fig:FigC}$a. While $\bar{\kappa}$ remains significant even for low $R_{\rm{sc}}$ in the RF MOT, it falls dramatically with decreasing $R_{\rm{sc}}$ in the DC* MOT.  The spread in $\bar{\kappa}$ for different $f_{\rm{MOT}}$ at low $R_{\rm{sc}}$ is typical of  day-to-day fluctuations in $\bar{\kappa}$  due to drifting alignment.  In the analytical model detailed in the Supplemental Material \cite{SMgeneral}, $\bar{\kappa}$ is related to $R_{\rm{sc}}$ by
\begin{equation}
\bar{\kappa} = -\frac{8 \Delta k R_{\rm{sc}} (1-2R_{\rm{sc}}/\Gamma_{\rm{eff}})\mu_{\rm{eff}} (\partial B_z/\partial z)}{3\times 4^{1/3} \Gamma^{2}(1+4\Delta^{2}/\Gamma^{2})},
\label{eq:kappa}
\end{equation}
where $k$ is the wavenumber, $\partial B_z/\partial z$ is the axial $B$-field gradient, and $\mu_{\rm{eff}}$ is an effective magnetic moment. The solid line in Fig.\,3a shows a fit of Eq.\,(\ref{eq:kappa}) to the data with $f_{\rm{MOT}} = 1.23$\,MHz. In this fit, we fix $\Gamma_{\rm{eff}}$ to the value found above and use experimental values \mbox{$\partial B_z/\partial z=9$\,G/cm} and $\Delta$\,=\,$-2\pi$\,$\times$\,9\,MHz, leaving $\mu_{\rm{eff}}$ as the only free parameter. The best fit gives $\mu_{\rm{eff}}$\,=\,0.32(1)\,$\mu_{B}$. This is not too different from the level-averaged value $\langle |\mu| \rangle = \sum_{i}^{n}|\mu_{i}|/n=0.49$\,$\mu_{B}$, where $\mu_{i}$ is the magnetic moment in the weak-field limit of Zeeman sub-level $i$, and the sum is over the sub-levels of ($v$\,=\,0,$N$\,=\,1). We see that the trapping forces in the RF MOT are reasonably well described by this simple model.

The geometric mean temperature $\bar{T}\equiv T_{\rho}^{2/3}T_z^{1/3}$ is shown vs $R_{\rm{sc}}$ in Fig.\,3b.  There is little frequency dependence to $\bar{T}$ for $f_{\rm{MOT}} > 1$\,MHz, while the 111\,kHz AC MOT is substantially hotter. The temperature decreases with decreasing scattering rate for all $f_{\rm{MOT}}$, qualitatively matching the expected behavior for Doppler cooling \cite{SMgeneral}:
\begin{equation}
T = -\frac{\hbar \Gamma^{2}}{8 k_{B} \Delta} \frac{(1+4\Delta^{2}/\Gamma^{2})}{(1-2R_{\rm{sc}}/\Gamma_{\rm{eff}})}.
\label{eq:DopplerTemperature}
\end{equation}
The solid line in Fig.\,3b shows this predicted behavior, with $\Gamma_{\rm{eff}}$ and $\Delta$ fixed as above.  There are no free parameters. At low $R_{\rm{sc}}$, we measure $\bar{T}$ as low as 400\,$\mu$K, fairly close to the 245\,$\mu$K predicted by Eq.\,\ref{eq:DopplerTemperature} (for SrF, the minimum Doppler cooling temperature $T_{\rm{D}}$\,=\,160\,$\mu$K would be expected for $\Delta$\,=\,$-\Gamma/2$\,$\approx$\,$-2\pi$$\times$3\,MHz).  Cooling to near the temperature predicted by simple Doppler cooling theory at low intensities has been observed in systems where sub-Doppler cooling mechanisms are weak or absent \cite{Xu2003,Chang2014}).  However, we observe $\bar{T}$ increases far more rapidly with increasing $R_{\rm{sc}}$ than suggested by Eq.\,(\ref{eq:DopplerTemperature}).  Similar behavior has been observed in Sr \cite{Xu2003} and is not well understood, but may be due to stimulated-force heating effects \cite{Gordon1980,Dalibard1985,Gould1991}.

Fig.\,\ref{fig:FigC}c shows the phase space density $\rho$ vs $R_{\rm{sc}}$, inferred from $\bar{\kappa}$, $\bar{T}$, and molecule number.  At the highest scattering rates, $\rho$ is similar for the DC, DC*, and RF MOTs.  Lowering $R_{\rm{sc}}$ increases $\rho$ by $\sim$2.5 orders of magnitude in the RF MOT, with $\rho$ roughly constant over the range $R_{\rm{sc}} \sim 10^5$\,-\,$10^6$\,s$^{-1}$.  The hotter $f_{\rm{MOT}}=111$\,kHz AC MOT has $\rho$\,$\sim$\,10$\times$ lower than RF MOTs with $f_{\rm{MOT}}$\,$>$\,1\,MHz.  In the DC* MOT, $\rho$ increases only modestly at lower $R_{\rm{sc}}$.

%Separate Loading and Trapping Stages

% Outlook:
The DC and DC* MOTs have polarization schemes intended to repump dark states with orthogonal and anti-confining MOT laser beams \cite{McCarron2015}, while in the RF MOT the laser polarizations and $B$-field are reversed before this repumping occurs.  The RF MOT was therefore expected to have a larger value of $\bar{\kappa}$ than the DC* MOT.  However, we find them to be comparable at high values of $R_{\rm{sc}}$ (Fig.\,\ref{fig:FigC}a).  Recently, it was shown that the close spacing of the SR/HF levels in SrF leads to a dual-frequency mechanism, where two oppositely-polarized frequency components addressing the same transition can deplete dark states \cite{TarbuttII2015,SMfigure}.  A state which is dark to one component is bright to the oppositely-polarized component, so molecules can scatter continuously from a confining beam, with a preference for that beam dictated by the Zeeman shift.  This mechanism is responsible for the majority of the confinement in the DC* MOT \cite{TarbuttII2015}.  In the RF MOT, dark states in all SR/HF levels are automatically converted to bright states by the time-varying fields.

While the RF MOT was not found to provide a larger restoring force than the DC* MOT, its ability to provide significant confinement even at greatly reduced power is extremely useful.  By loading at high power and ramping to low power, we produce a sample of 2000 molecules with 0.5\,s lifetime at 400\,$\mu$K with density $6\times10^{4}$\,cm$^{-3}$ and phase space density $1.5\times10^{-14}$.  These correspond to factors of 3, 4, 6, 15, and 1000 improvement, respectively, over previous molecular MOTs \cite{Barry2014,McCarron2015}.  There are promising ideas for delivering much larger numbers of slow molecules for loading into an RF MOT, such as microwave guiding \cite{DeMille2013}, bichromatic force slowing \cite{Chieda2011}, or (for molecules with a $^1\Sigma$ ground state, where the magnetic moment is negligible) a Zeeman slower \cite{Hendricks2014,Barry2013}.

The temperatures reported here, $\approx$\,3\,$T_{\rm{D}}$, are sufficiently cold to have a broad impact on the field.  For example, molecular fountains could be constructed to offer long interaction times for precision measurements \cite{Tarbutt2013}.  Additionally, 400\,$\mu$K molecules could be magnetically trapped with standard techniques.  Co-trapping with an ultracold atom would allow study of atom-molecule collisions and potentially sympathetic cooling to yet lower temperatures \cite{Lim2015,Lara2006,Soldan2009}.

The authors acknowledge E R Edwards and N R Hutzler for input on RF MOT coil design and financial support from ARO and ARO (MURI). EBN acknowledges funding from the NSF GRFP.

\bibliography{thebib}

\clearpage

\section{SUPPLEMENTAL MATERIAL}

\subsection{Polarization and Frequency Schemes}
\begin{figure*}[!t]
\centering
\includegraphics[width= .8\linewidth]{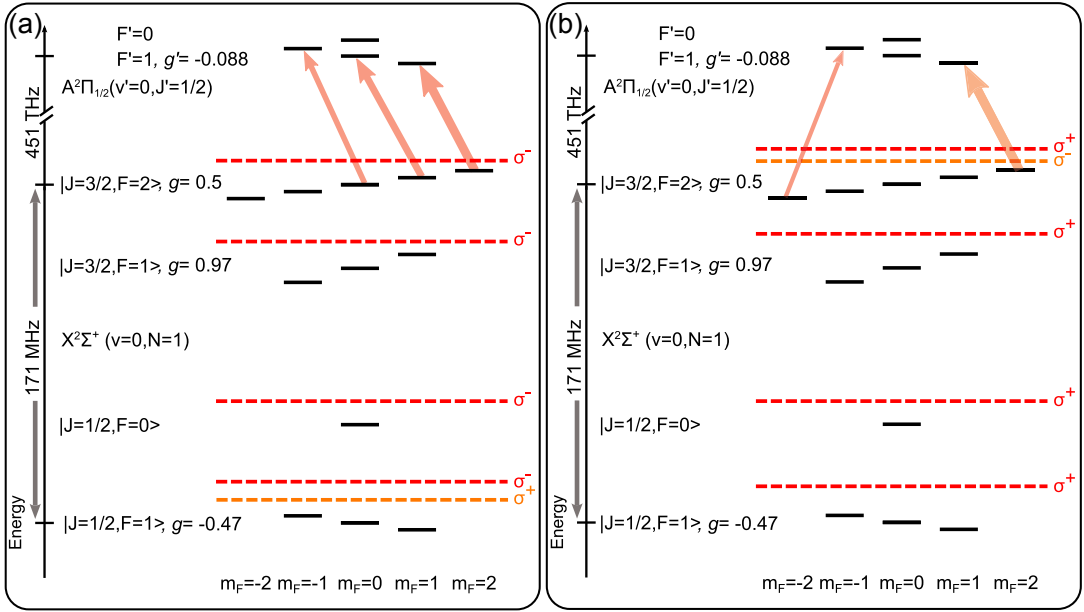}
        \caption{Polarization scheme for (a) RF and DC MOTs, and (b) DC* MOT.  Dashed lines show the ground-state energy with which $\mathcal{L}_{00}^{\dagger}$ (orange) and each $\mathcal{L}_{00}$ frequency component (red) would be resonant for the set polarization.  Excited state hyperfine components are not resolved (splitting less than $\Gamma$). In (b), arrows illustrate repumping via the dual-frequency effect.
Zeeman shifts and polarizations are shown for $B$\,$>$\,0, with excited state Zeeman shifts and hyperfine splittings exaggerated for clarity.}\label{fig:FigE}
\end{figure*}

The ground state of the main cycling transition in SrF has a manifold of four SR/HF levels, each of which must be addressed by an individual optical frequency for significant confinement. In both the RF and DC MOTs, the polarizations of individual laser frequency components are chosen so that the ground states Zeeman-shifted closest to resonance are bright states (Fig.\,\ref{fig:FigE}a).  In fact, for a MOT with static polarizations where each transition is driven by a single frequency component, the polarization preference is dictated by the change in total angular momentum $\Delta F$, and by the Zeeman splitting of the excited state (but not of the ground state) \cite{Tarbutt2015}. In the DC* MOT, the polarization of each frequency component is chosen according to that principle (Fig.\,\ref{fig:FigE}b). In this way, the bright states are preferentially excited by the confining laser beam, while population is primarily pumped out of dark states by orthogonal beams rather than by the anti-confining beam.  In the RF MOT the laser polarizations and $B$-field are reversed before significant repumping from dark states by non-confining lasers occurs, so a large confining force is expected as molecules continuously scatter photons from the confining beam.  In the DC* MOT, a dual-frequency mechanism \cite{TarbuttII2015} (Fig.\,\ref{fig:FigE}b) depletes dark states faster than non-confining beams and provides the majority of the confinement.

\subsection{Simple model of the MOT}

A simple rate model of the multi-level molecule [4] considers $n_g$ ground states excited to $n_e$ excited states, and finds that the steady-state scattering rate is
\begin{equation}
R_{\rm{sc}} = \Gamma \frac{n_e}{(n_g + n_e) + 2\sum_{j=1}^{n_{g}}(1+4\Delta_{j}^{2}/\Gamma^{2})I_{\rm{sat},j}/I_{j}}.
\label{eq:Rsc1}
\end{equation}
Here, $\Gamma$\,=\,$2\pi$$\times$7\,MHz is the spontaneous decay rate of the transition, $I_j$ is the intensity of the light driving the transition from ground state $j$, $I_{\rm{sat},j}=\pi h c \Gamma/(3 \lambda_{j}^{3})$ is the usual two-level saturation intensity for a transition of wavelength $\lambda_{j}$, $\Delta_j$ is the detuning of the light from the resonance frequency, and the sum runs over all ground states. To apply this model to our experiment, we include the 24 levels of ($v$\,=\,0, $N$\,=\,1) and ($v$\,=\,1, $N$\,=\,1), all of which are coupled to the same four levels of the excited state. In addition, we include the microwave coupling between $N$\,=\,0 and $N$\,=\,1, which effectively couples the four $N$\,=\,0 levels to the same excited state. Other vibrational and rotational states involved in the experiment can safely be neglected because the population in these states is negligible when the repump lasers are on. So we have $n_e=4$, $n_g=28$ with microwave coupling and $n_g=24$ without. The lasers addressing the $v=1$ levels have $\Delta\simeq 0$, whereas those addressing $v=0$ have $\Delta \simeq -1.4\Gamma$. Furthermore, the intensity of the ${\cal L}_{10}$ laser exceeds that of ${\cal L}_{00}$ over the whole range of powers explored. Therefore, the $v=1$ transitions contribute less than 10\% to the sum over transitions in Eq.\,(\ref{eq:Rsc1}), and we neglect them. The total intensity $I_{00}$ driving the $n_{g0}=12$ transitions from $v=0$ is divided roughly equally between them, so we write $I_{j}=I_{00}/n_{g0}$. Furthermore, because all transitions are excited by nearly identical wavelength light with similar detunings, we set $I_{\rm{sat},j}$\,=\,$I_{\rm{sat}}$ and $\Delta_j$\,=\,$\Delta$ for all $j$. This gives

\begin{equation}
R_{\rm{sc}} = \Gamma \frac{n_{e}}{(n_g + n_e) + 2(1+4\Delta^{2}/\Gamma^{2})n_{g0}^{2}I_{\rm{sat}}/I_{00}},
\end{equation}
which we can re-arrange to the familiar form
\begin{eqnarray}
\label{eq:Rsc2}
&R_{\rm{sc}} = \frac{\Gamma_{\rm{eff}}}{2} \frac{s_{\rm{eff}}}{1 + s_{\rm{eff}} + 4\Delta^{2}/\Gamma^{2}}\\
\label{eq:GammaEff}
&\Gamma_{\rm{eff}} = \frac{2 n_{e}}{n_{g} + n_{e}} \Gamma,\\
\label{eq:sEff}
&s_{\rm_{eff}} =  \frac{I_{00}}{I_{\rm{sat,eff}}} = \frac{(n_{g} + n_{e})}{2 n_{g0}^{2}} \frac{I_{00}}{I_{\rm{sat}}},
\end{eqnarray}
where $s_{\rm{eff}}$ is an effective saturation parameter.

The scattering force due to a single beam is $\hbar k R_{\rm{sc}}$, where $k=2\pi/\lambda$ is the wavenumber. To find the spring constant of the MOT, we sum the forces due to a pair of counter-propagating beams at an axial position $z$ where the Zeeman shift is $\mu_{\rm{eff}} z (\partial B_z/\partial z)$. Here, $\mu_{\rm{eff}}$ is the effective magnetic moment and $\partial B_z/\partial z$ is the axial magnetic field gradient. The axial gradient of this force, evaluated at the origin, is the axial spring constant,
\begin{equation}
\kappa_z = -\frac{8 \Delta k \Gamma_{\rm{eff}} \mu_{\rm{eff}}  (s_{\rm{eff}}/6)(\partial B_z/\partial z)}{\Gamma^{2}(1+s_{\rm{eff}}+4\Delta^{2}/\Gamma^{2})^{2}}.
\end{equation}
The approach of summing the forces due to individual beams does not correctly account for saturation. Following a common ad hoc approach \cite{Lett1989}, we have used the full intensity ($s_{\rm{eff}}$) seen by the molecule in the saturation term that appears in the denominator, while the intensity that appears in the numerator is the single beam intensity ($s_{\rm{eff}}/6$). In this form, the expression has been found to work quite well for modest saturation parameters \cite{Lett1989}, even though it is strictly only valid in the limit of low saturation. Using Eq.\,(\ref{eq:Rsc2}), we express this result in terms of the scattering rate:
\begin{equation}
\kappa_z = -\frac{8 \Delta k R_{\rm{sc}} (1-2R_{\rm{sc}}/\Gamma_{\rm{eff}})\mu_{\rm{eff}}(\partial B_z/\partial z)}{3 \Gamma^{2}(1+4\Delta^{2}/\Gamma^{2})}.
\end{equation}
The geometric mean spring constant is $\bar{\kappa} = \kappa_{z}/4^{1/3}$.

In a simple theory of the MOT, the equilibrium temperature is found by equating the heating rate due to the randomness of photon scattering with the Doppler cooling rate. Both heating and cooling rates are proportional to the scattering rate, and so the multi-level molecule should have the same temperature as a simple two-level system:
\begin{equation}
T = -\frac{\hbar \Gamma^{2}}{8 k_{B} \Delta}(1 + s_{\rm{eff}} + 4\Delta^{2}/\Gamma^{2}).
\end{equation}
Expressing this in terms of the scattering rate we obtain
\begin{equation}
T = -\frac{\hbar \Gamma^{2}}{8 k_{B} \Delta} \frac{(1+4\Delta^{2}/\Gamma^{2})}{(1-2R_{\rm{sc}}/\Gamma_{\rm{eff}})}.
\end{equation}

To connect most easily with the experiment, we re-write Eq.\,(\ref{eq:Rsc2}) as
\begin{equation}
R_{\rm{sc}} = \frac{\Gamma_{\rm{eff}}}{2} \frac{p_{00}/p_{\rm{sat}}}{1 + p_{00}/p_{\rm{sat}} + 4\Delta^{2}/\Gamma^{2}},
\label{eq:Rsc3}
\end{equation}
where $p_{00}$ is the single-beam power of ${\cal L}_{00}$, and $p_{\rm{sat}}$ is an effective saturation power given by
\begin{equation}
p_{\rm{sat}} = \frac{1}{6\times 1.35} \frac{\pi w_{0}^{2}}{2} \frac{2 n_{g0}^{2}}{(n_{g} + n_{e})} I_{\rm{sat}}.
\label{eq:pSat}
\end{equation}
Here, we have assumed that the molecules are at the center of the MOT where the intensity due to all 6 beams of ${\cal L}_{00}$ is $I_{00}$\,=\,$6 p_{00}/(\pi w_{0}^{2}/2)$, where $w_{0} = 7$\,mm is the $1/e^2$ radius of each beam. The extra factor of 1.35 in Eq.\,(\ref{eq:pSat}) accounts for the additional power due to ${\cal L}_{00}^{\dagger}$ which was always 35\% of $p_{00}$ in the experiment. Our model (including the microwave coupling) gives $\Gamma_{\rm{eff}}$\,=\,$\Gamma/4$ and $p_{\rm{sat}}$\,=\,2.6\,mW.

\subsection{Multilevel Rate Model}

In addition to this simple analytical model, we performed multilevel simulations using the rate model described in \cite{Tarbutt2015}, refined to account for the non-linear Zeeman shifts of the lower levels, the microwave coupling to $N$\,=\,0, the sinusoidal oscillation of the magnetic field, and the square-wave modulation of the polarizations with a variable phase difference. At each zero crossing of the magnetic field the population in hyperfine component $(F,m)$ is switched to component $(F,-m)$. The model includes lasers ${\cal L}_{00}$, ${\cal L}_{00}^{\dagger}$ and ${\cal L}_{10}$, neglecting repumping from other negligibly populated vibrational and rotational states via transitions not coupled to the main cycling transition. The simulation is used to predict $R_{\rm{sc}}$, $\kappa$, and damping constant $\alpha$, and the temperature is inferred from the relation $k_B T$\,=\,$(\hbar k)^{2}R_{\rm{sc}}/\alpha$.  We find that the analytical model discussed above captures most of the properties predicted in the more detailed multilevel model, justifying fitting the experimental data to this simpler model.

Figure \ref{modelVsSwitchFreq} shows the simulated axial spring constant and damping rate as a function of $f_{\rm{MOT}}$ for a set of isolated angular momentum cases relevant to the experiment. In each case, the system consists of a lower level with angular momentum $F$ and an upper level with angular momentum $F'$, addressed by six 20\,mW MOT beams containing a single frequency component with polarization as in the experiment (see Fig.\,\ref{fig:FigE}). All the parameters, including the $g$-factors, are the same as for the SrF system. In all cases, we find that $R_{\rm{sc}}$ is nearly independent of $f_{\rm{MOT}}$, whereas the spring constant and damping rate have a strong dependence. Hence within this model, the enhanced force in the RF MOT is due to an increase in the proportion of photons scattered from the confining beam, rather than an increase in the total scattering rate. In this model there is also a clear transition from AC to RF MOT operation: at low frequencies both $\kappa$ and $\alpha$ are independent of $f_{\rm{MOT}}$ and have values that are characteristic of a DC MOT, while at high frequencies, where $f_{\rm{MOT}} \sim R_{\rm{sc}}$, they are again independent of $f_{\rm{MOT}}$ with higher values characteristic of the RF MOT.

In the case where $F$\,=\,0, the RF MOT is not helpful because the transition to $F'$\,=\,1 can always be driven by either polarization, and the MOT behaves like the DC case.  Because the excited state $g$-factor is negative, the correct choice of polarization is opposite to the one used in the experiment. Consequently, the spring constant is negative (we have plotted $-\kappa$ so that we can show the result on the log scale). This is observed experimentally (Fig.\,\ref{modelVsSwitchFreq}c): compared to when oppositely polarized laser $\mathcal{L}_{00}^{\dagger}$ is off resonant with all transitions, the LIF signal from the MOT is observed to increase slightly when $\mathcal{L}_{00}^{\dagger}$ is detuned to the red of the $F$\,=\,0\,$\rightarrow$\,$F'$\,=\,1 transitions, but much less than when tuned as in Fig.\,\ref{fig:FigE}a. The polarization scheme is therefore suboptimal for the overall confining force, though only slightly since only a small fraction of the total population is in the $F$\,=\,0 state.

For the other three possible transitions from the ground state manifold, the simulations predict that $\kappa$ and $\alpha$ increase with increasing $f_{\rm{MOT}}$.  The simulated spring constant increases by about an order of magnitude for the $F=1\rightarrow F'=1$ and $F=2\rightarrow F'=1$ transitions from the AC MOT to the RF MOT.  For the $F=1\rightarrow F'=0$ transition, the increase is even more dramatic as the optimal polarization is reversed for DC and RF operation \cite{Tarbutt2015}.  The simulated damping does not change so dramatically with $f_{\rm{MOT}}$, but in all cases the RF MOT has greater damping than the AC MOT.

\begin{figure}[!t]
 \centering
 \includegraphics[width=0.38\textwidth]{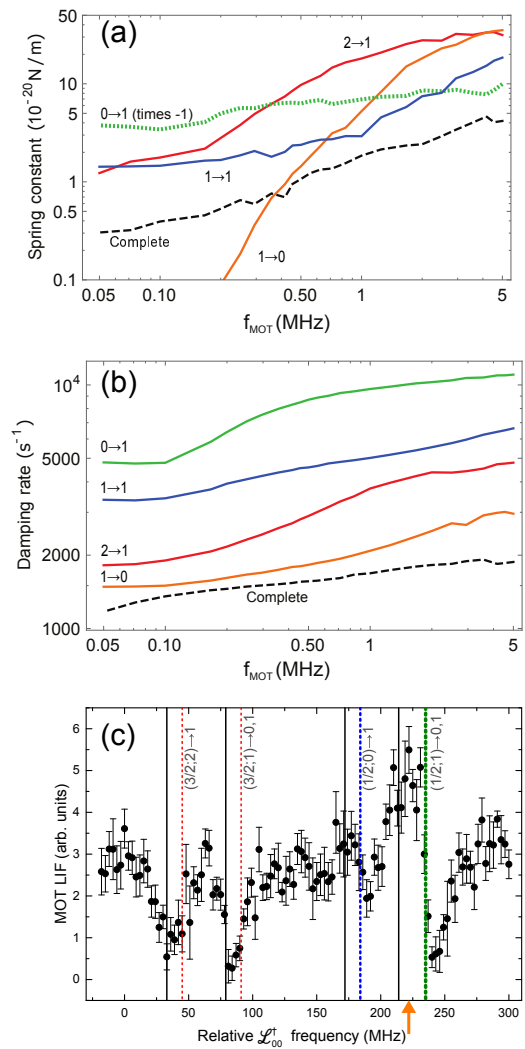}
\caption{\label{modelVsSwitchFreq} Predictions from the multilevel rate model for \mbox{(a) axial spring} constant $\kappa_z$ and (b) damping rate $\alpha$ as a function of $f_{\rm{MOT}}$ for a set of simple angular momentum cases relevant to the experiment (solid and short-dashed lines), and a complete multilevel simulation for SrF (long-dashed lines).  The individual cases are labeled as $F$\,$\rightarrow$\,$F'$, where $F$ and $F'$ are the angular momenta of the lower and upper states. Note that for 0\,$\rightarrow$\,1 (short-dashed line) we have plotted $-\kappa_z$ in (a). (c) Experimental data showing LIF signal in RF MOT vs $\mathcal{L}_{00}^{\dagger}$ frequency. Dashed lines denote resonance with the labeled $(J;F)$\,$\rightarrow$\,$F'$ transition, while solid lines denote the frequencies of the four RF sidebands of $\mathcal{L}_{00}$.  Significantly more LIF is observed when $\mathcal{L}_{00}^{\dagger}$ is slightly red-detuned from the  \mbox{$(1/2;1)$\,$\rightarrow$\,$1,0$} (green, highest frequency) transition, and a slight increase in LIF is observed when $\mathcal{L}_{00}^{\dagger}$ is red-detuned from the \mbox{$(1/2;0)$\,$\rightarrow$\,$1$} (blue, second highest frequency) transition.  Reduced LIF for $\mathcal{L}_{00}^{\dagger}$ blue-detuned from these transitions is attributed to Doppler heating; tuning near the two lower frequency transitions (red) weakens the confinement and lowers the LIF signal.  The orange arrow shows the $\mathcal{L}_{00}^{\dagger}$ frequency during normal RF MOT operation.  Error bars are standard errors of multiple scans.
}\label{modelVsSwitchFreq}
\end{figure}

Figure \ref{modelVsSwitchFreq} also shows the model results for the complete SrF system, including the microwave coupling and repumping by ${\cal L}_{10}$. All parameters are as in the experiment, and we have taken the ${\cal L}_{00}$ power to be 80\,mW, leading to a scattering rate $R_{\rm{sc}}$\,$\approx$\,4.6$\times$$10^6$\,s$^{-1}$.  The model predicts that at full laser power, the spring constant increases until $f_{\rm{MOT}} \simeq R_{\rm{sc}}$, with $\kappa$\,$\sim$\,10$\times$ larger in the RF MOT than in the AC MOT.  In the experiment, however, we observe that $\bar{\kappa}$ is roughly independent of $f_{\rm{MOT}}$, even in the 110 kHz AC MOT.  The model also predicts that $\alpha$ does not depend strongly on $f_{\rm{MOT}}$.  However, in our experiments the 110 kHz AC MOT temperature is observed to be consistently higher than for $f_{\rm{MOT}}$\,$>$\,1\,MHz.  Hence it appears that this multilevel model is not able to reproduce some striking aspects of the experimental data.
\begin{figure}[!b]
 \centering
 \includegraphics[width=0.5\textwidth]{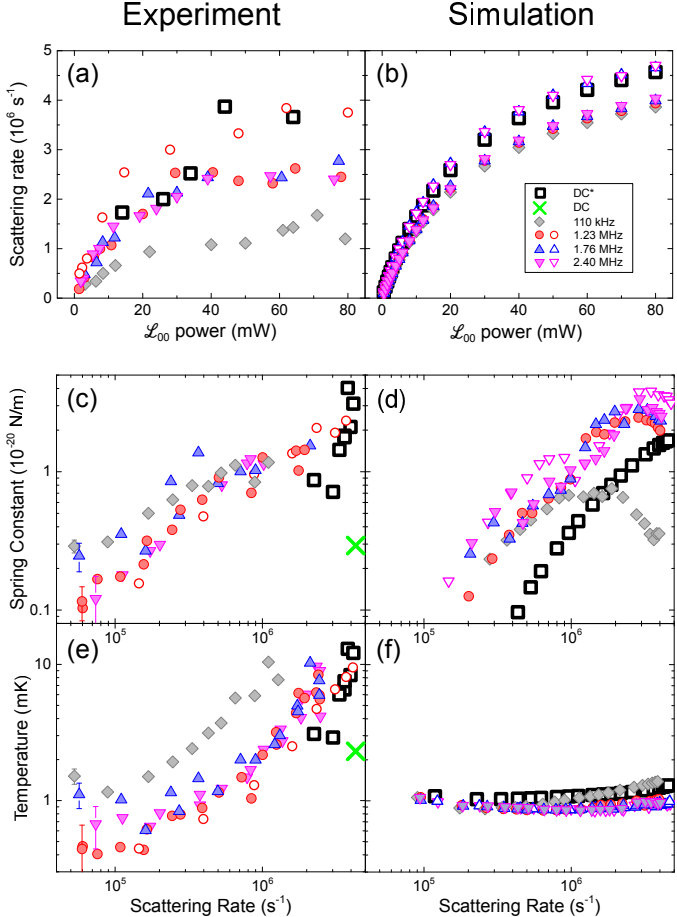}
\caption{(Marker legend in figure.  Filled markers with microwaves applied, open markers without.)  Comparison of trap properties measured in experiment (left column) and calculated from the multilevel simulation (right column).  (a,\,b) $R_{\rm{sc}}$ vs $\mathcal{L}_{00}$ power.  Observed scattering rates are consistent with the model for $f_{\rm{MOT}}$\,$>$\,1\,MHz, but $\sim$2$\times$ lower for \mbox{$f_{\rm{MOT}}$\,=\,110 kHz}. Microwaves reduce $R_{\rm{sc}}$ by roughly the expected amount.  (c,\,d)  $\bar{\kappa}$ vs $R_{\rm{sc}}$.  The measured $\bar{\kappa}$ is consistent with the model for low $R_{\rm{sc}}$, and $\sim$2$\times$ weaker than calculated at full power.  $\bar{\kappa}$ is observed to be $\sim$2$\times$ larger for the DC* MOT than the RF MOT at full power, while the simulation predicts $\bar{\kappa}$ slightly larger in the DC* MOT.  (e,\,f) $\bar{T}$ vs $R_{\rm{sc}}$.  $\bar{T}$ decreases with $R_{\rm{sc}}$ to a minimum value.  $\bar{T}$ is much lower for $f_{\rm{MOT}}$\,$>$\,1\,MHz than for $f_{\rm{MOT}}$\,=\,110 kHz.  The simulation does not capture any of the measured temperature behavior aside from the correct order of magnitude.  For (c) and (e), the spread in values for a given MOT configuration are indicative of variations between measurements made on the same day, while the representative error bars are standard errors of measurements taken on multiple days under similar conditions.
}\label{fig:Fig10}
\end{figure}

In Fig.\,6, we compare data presented in the main body of the paper to the multilevel simulation.  The measured scattering rate (Fig.\,6a) agrees reasonably well with the model prediction (Fig.\,6b), except $R_{\rm{sc}}$ is found to be $\sim$2$\times$ lower than predicted in the $f_{\rm{MOT}}=111$ kHz AC MOT.  When $N$\,=\,0 and $N$\,=\,1 are coupled by microwaves, increasing $n_g$ from 24 to 28, Eq.\,\ref{eq:Rsc2} predicts $R_{\rm{sc}}$ to be 13\,\% smaller than the case without microwaves present; we measure a slightly larger $\approx$40\,\% decrease.

The measured spring constant $\bar{\kappa}$  (Fig.\,6c) increases proportional to $R_{sc}$ until saturating at a maximum value.  The simulation (Fig.\,6d) agrees well with the experiment for $R_{\rm{sc}}$\,$\lesssim$\,10$^6$\,s$^{-1}$. At higher scattering rate it slightly underestimates $\bar{\kappa}$ for $f_{\rm{MOT}}$\,=\,111\,kHz, but overestimates $\bar{\kappa}$ for $f_{\rm{MOT}}$\,$>$\,1\,MHz by $\sim$2$\times$.   The simulation also correctly predicts a marginally higher confining force in the absence of microwaves at full laser power.   Contrary to the simulation, the RF MOT is found to provide a lower restoring force than the DC* MOT at full power.  Confinement in the DC* MOT is due primarily to the dual frequency mechanism; the spring constant is reduced by a factor of 5 when these effects are removed in the model.  As the $R_{\rm{sc}}$ is lowered, we find that the spring constant of the RF MOT drops as predicted both by our simple analytical model and our detailed numerical model. The measured spring constant of the DC* MOT decreases far more rapidly, and this observation is not yet explained.

The measured temperature $\bar{T}$ (Fig.\,6e) decreases linearly with decreasing scattering rate for all $f_{\rm{MOT}}$, approaching an intercept as low as $\sim$400 $\mu$K, close to the Doppler cooling limit.  The simulation (Fig.\,6f) predicts a temperature nearly independent of trap configuration. This suggests that, at high scattering rate, there is either some additional heating mechanism or that the damping is slower than expected.  The ability of the RF MOT to provide sufficient confinement at greatly reduced power has been the key to achieving low temperature and high phase space density.  As neither the analytical model or rate equation model predict the rapid increase in temperature with increasing scattering rate observed in the experiment, the high temperature must be due physics not captured in either model.

To test the validity of fitting the experimental data to the simple analytical model rather than the multilevel simulations, we have fit the multilevel simulations to the analytical model and find they are in good agreement.  With the experimental value $\Delta$\,=\,-2$\pi$$\times$9\,MHz, the best fit parameters to the simulation data for Eq.\,(13) are $\Gamma_{\rm{eff}}$\,=\,$\Gamma/4$, exactly as expected from Eq.\,(7), and $p_{\rm{sat}}$\,=\,3.6\,mW, which is about 40\% higher than predicted by Eq.\,(14). With $\Gamma_{\rm{eff}}$ now fixed, we fit Eq.\,(2) to the simulation data in Fig.\,6d. This fit gives the best fit parameter $\mu_{\rm{eff}}$\,=\,0.36\,$\mu_{B}$, which is close to the value 0.32\,$\mu_{B}$ found by fitting to the experimental data. Finally, the temperature found from the model is about twice that predicted by Eq.\,(3), except at high $R_{\rm{sc}}$ where they are closer. Note that here we have fixed $\Gamma_{\rm{eff}}$\,=\,$\Gamma/4$ as found above and that this is 75\% higher than found in the experiment. This difference explains the difference in form between the simulated data in Fig.\,6f and the theory curve in Fig.\,3b.

\subsection{DC Magnetic Fields}

	We have considered the effects of a DC component of the magnetic field $B_{\rm{DC}}$ on the RF MOT.  One issue is that the field zero will oscillate along the direction of $B_{\rm{DC}}$ at $f_{\rm{MOT}}$, which could induce heating.  Additionally, in a coordinate system whose $z$-axis is along the local $B$-field, the RF MOT relies on diabatic transfer of molecules from states with total angular momentum projection $m$ to $- m$ to destabilize dark states.  How rapidly the field must be switched depends upon the zero-field energy splitting of these two states.  These states are degenerate with no applied field and the state transfer probability is unity.  However, a stray, DC component of the magnetic field can lead to avoided crossings.  A simple two-level Landau-Zener model can approximate the probability $\mathbb{P}$ of successful state transfer:
\begin{equation}
\mathbb{P}~=~e^{-4 g \mu_B B_{\rm{DC}}^2/f_{\rm{MOT}} B_{\rm{pk}}},
\end{equation}
where $\mu_B$ is the Bohr magneton and $B_{\rm{pk}}$ is the peak value of the RF $B$-field.  If $\mathbb{P}$ is too low, the RF MOT mechanism will be ineffective and dark states will destabilize at a much slower rate via DC MOT mechanisms.

As an estimate of an acceptable value for $B_{\rm{DC}}$, \mbox{$\mathbb{P}$\,$<$\,0.5} when \mbox{$B_{\rm{DC}}$\,$>$\,0.3\,G} for typical conditions $B_{\rm{pk}}$\,=\,4\,G, $f_{\rm{MOT}}$\,=\,1\,MHz, and $g$\,=\,1.  However, this model is exponentially sensitive to $g$, which differs for each SR/HF level, and this constraint on $\mathbb{P}$ may be too conservative.  In practice, no significant change in LIF intensity, $\tau_{\rm{MOT}}$, $\bar{T}$, or $\bar{\kappa}$ was observed experimentally for applied $B_{\rm{DC}}$ up to 1\,G.  When slowing, we apply a $\sim$3.5\,G field to remix dark states via Larmor precession.  This field is canceled in the MOT region by oppositely wound coils during the 35 ms of slowing.  Both the remixing and cancellation coils are turned off in $\sim$100\,$\mu$s after the slowing period, and no DC fields are actively applied during normal trapping.

\subsection{RF Electric Fields}
 As described in the main text, the large voltage required to drive the MOT coils at radio-frequencies can lead to an unwanted $\mathcal{E}$-field ($\mathcal{E}$ fields due to the time dependence of $B$, i.e. Faraday's Law, should be negligibly small, $\sim$10\,mV/cm).  This $\mathcal{E}$-field in turn leads to mixing of opposite-parity states in the molecule, including (most significantly) the close-lying $\Lambda$-doublet pair in the excited state of our main cycling transition.  This mixing leads to unwanted decays to rotational states that are not part of the optical cycle.

 Originally, the MOT coils were wired in series with the two coils on the bottom AlN board preceding the two on the top board, leading to root-mean-square fields $\mathcal{E}_{\rm{RF}}$\,$\sim$\,100\,V/cm.  In this arrangement, with $f_{\rm{MOT}}=1.9$\,MHz, we observed lifetimes as short as $\tau_{\rm{MOT}}=19$\,ms without microwaves and $\mathcal{L}_{00}^{N=2}$ applied to provide remixing from the rotational states populated by $\mathcal{E}$-induced decays.  This is significantly longer than the $\sim$1\,ms lifetime expected from the known $\Lambda$-doublet splitting \cite{Sheridan2009}, electric dipole matrix element \cite{Kandler1989}, and measured $R_{\rm{sc}}$.  Rearranging the coils to the configuration described in the main text decreased $\mathcal{E}_{\rm{RF}}$ in the trap center by $\sim$$5\times$,  as measured using a home-built Schottky diode-based RF $\mathcal{E}$-field probe.  With the rewired coils, we observed $\sim$$5\times$ slower loss into $N=0,2$, though the loss rate is expected to scale as the square of $\mathcal{E}$.

 There exist coil geometries which better cancel the stray field from the effective capacitor formed by the coil traces on either side of the AlN board than our simple spiral pattern.  Using the same $\mathcal{E}$-field probe, we find that $\mathcal{E}_{\rm{RF}}$ may be reduced by an additional $\sim$$50\times$ using such alternative geometries.  Such low fields would presumably eliminate the need for the repump laser $\mathcal{L}_{00}^{N=2}$ and for the remixing microwaves.  These measurements are in reasonable agreement with simulations performed with finite element software.

\subsection{Lifetime}
By varying the background pressure and He buffer gas flow, we verified that collisions with background gas and ballistic He from our source are not the dominant loss mechanisms, even at the longest lifetimes observed.  The cause of the observed scattering-independent loss (Fig.\,\ref{fig:FigD}c) will be investigated in future work.

\end{document}